\documentclass[aps,prl,twocolumn,groupedaddress,amsmath,amssymb,floatfix]{revtex4-1}
\usepackage{graphicx}  
\usepackage{verbatim}   
\usepackage{enumerate}
\usepackage{mathtools}
\usepackage{xcolor}
\makeatletter
\newcommand{\vast}{\bBigg@{3}}
\newcommand{\Vast}{\bBigg@{5}}

\usepackage[colorlinks=true,linkcolor=red,citecolor=blue]{hyperref}
\usepackage{amsthm,amsmath,amssymb,bm,bbm,yhmath,mathrsfs,float}
\DeclareFontFamily{U}{mathx}{\hyphenchar\font45}
\DeclareFontShape{U}{mathx}{m}{n}{
<5> <6> <7> <8> <9> <10>
<10.95> <12> <14.4> <17.28> <20.74> <24.88>
mathx10}{}
\DeclareSymbolFont{mathx}{U}{mathx}{m}{n}
\DeclareFontSubstitution{U}{mathx}{m}{n}
\DeclareMathAccent{\widecheck}{0}{mathx}{"71}
\def\ujp{|u|}
\def\Gag{\Ga_g}
\def\Gab{\Ga_b}
\def\Gaw{\Ga_w}
\def\Gar{\Ga_r}

\def\nabo{\overset{\circ}{\nab}}
\def\divo{\overset{\circ}{\sdiv}\,}
\def\curlo{\overset{\circ}{\curl}\,}
\def\dodo{\overset{\circ}{d}}

\def\trchc{\widecheck{\trch}}
\def\trchbc{\widecheck{\trchb}}

\def\sk{\mathfrak{s}}

\def\D{\mathbf{D}}
\def\DD{\mathcal{D}}

\renewcommand{\c}{\cdot}

\DeclareMathOperator{\sdiv}{div}

\def\D{\mathbf{D}}

\def\th{\theta}
\def\deo{\overset{\circ}{\de}}
\def\debo{\overset{\circ}{\deb}}

\def\mm{{\overline{m}}}

\def\deb{{\overline{\de}}}

\newcommand{\bea}{\begin{eqnarray}}
\newcommand{\eea}{\end{eqnarray}}
\newcommand{\ov}{\overline}
\def\beaa{\begin{eqnarray*}}
\def\eeaa{\end{eqnarray*}}

\newtheorem{thm}{Theorem}
\newtheorem{prp}[thm]{Proposition}

\newtheorem{df}[thm]{Definition}

\newtheorem{prop}[thm]{Proposition}
\newtheorem{lem}[thm]{Lemma}

\def\Xh{{\widehat{X}}}
\def\Xhb{{\widehat{\underline{X}}}}

\def\om{\omega}

\def\curls{\curl}

\newcommand{\la}{\lambda}

\def\II{\mathscr{I}}

\allowdisplaybreaks[3]
\def\MM{\mathcal{M}}
\DeclareMathOperator{\curl}{curl}

\newcommand{\f}{\frac}

\DeclareMathOperator{\tr}{tr}

\def\The{\Theta}

\newcommand{\vphi}{\varphi}
\renewcommand{\a}{\alpha}
\renewcommand{\b}{\beta}
\newcommand{\Ga}{\Gamma}
\newcommand{\pr}{\partial}
\newcommand{\chib}{{\underline{\chi}}}
\newcommand{\etab}{{\underline{\eta}}}
\newcommand{\xib}{{\underline{\xi}}}
\newcommand{\omb}{{\underline{\omega}}}
\newcommand{\g}{{\bf g}}

\renewcommand{\aa}{\underline{\a}}

\newcommand{\bb}{\underline{\b}}

\DeclareMathOperator{\im}{Im}
\def\M{\MM}

\def\trch{\tr\chi}
\def\trchb{\tr\chib}

\def\Si{\Sigma}

\def\hch{{\widehat{\chi}}}
\def\hchb{{\widehat{\chib}}}

\def\ze{\zeta}
\def\nab{\nabla}
\def\De{\Delta}
\def\R{\mathbf{R}}

\def\JJ{\mathcal{J}}
\def\D{{\bf D}}
\def\de{\delta}
\def\si{\sigma}

\def\DD{\mathcal{D}}

\def\PP{\mathcal{P}}
\def\QQ{\mathcal{Q}}

\def\Ab{{\underline{A}}}
\def\Bb{{\underline{B}}}
\def\rhod{\si}
\def\Thb{{\underline{\Theta}}}
\def\s2{\sqrt{2}}
\DeclareMathOperator{\Ric}{{\bf Ric}}

\begin{document}
\title{Angular Momentum Memory Effect}
\author{Xinliang An}
\address{Department of Mathematics, National University of Singapore, Singapore 119076}
\author{Taoran He}
\address{Department of Mathematics, National University of Singapore, Singapore 119076}
\author{Dawei Shen}
\address{Laboratoire Jacques-Louis Lions, Sorbonne Universit\'e, 75252 Paris, France}
\begin{abstract}
\noindent Utilizing recent mathematical advances in proving stability of Minkowski spacetime with minimal decay rates and nonlinear stability of Kerr black holes with small angular momentum, we investigate the detailed asymptotic behaviors of gravitational waves generated in these spacetimes. Here we report and propose a new angular momentum memory effect along future null infinity. This accompanies Christodoulou's nonlinear displacement memory effect and the spin memory effect. The connections and differences to these effects are also addressed.
\end{abstract}
\maketitle
\emph{Introduction.---}Rapid progress has been made in the hyperbolic theory of mathematical general relativity. In particular, Kerr nonlinear stability with small angular momentum has been recently proven by Klainerman-Szeftel, Giorgi-Klainerman-Szeftel and the third author in the series of works \cite{KS:Kerr1,KS:Kerr2,KS:main,GKS,Shen}. In addition, the global stability of Minkowski spacetime, first revealed by Christodoulou-Klainerman \cite{Ch-Kl}, has been reproved under minimal decay assumptions of initial data by the third author \cite{Shenglobal}. Detailed explicit hyperbolic estimates have been provided. With these, we revisit the field of memory effects of gravitational waves. In particular, in this article, we extend Christodoulou's displacement memory \cite{Ch} and Bieri's extension \cite{BieriMemory} to broader settings. Employing the newly defined intrinsic \textit{angular momentum} introduced by Klainerman-Szeftel in \cite{KS:Kerr2}, we identify a new \textit{angular momentum memory effect}. Links and differences to the spin memory effect \cite{PSZ} are also discussed. In a later section, we also translate our results into Newman-Penrose (NP) formalism.

\vspace{2mm}

\emph{Preliminaries.---}In a $3+1$ dimensional Lorentzian manifold $(\M,\g)$, we study the Einstein vacuum equations:
\begin{equation}\label{EVE}
    \Ric_{\mu\nu}(\g)=0 \quad \mbox{with}\quad \mu,\nu\in\{1,2,3,4\}.
\end{equation}
We foliate the spacetime $\MM$ by \emph{maximal hypersurfaces} $\Si_t$ as level sets of a time function $t$ and by outgoing null cones $C_u$ as level sets of an \emph{optical function} $u$. The intersections of $\Si_t$ and $C_u$ are 2-spheres denoted by $ S(t,u)$ with $u\in\mathbb{R}, t\geq 0$. We define the \emph{area radius} of $S(t,u)$ by $r:=\sqrt{{|S(t,u)|}/{4\pi}}$. Let $T$ be the future-oriented unit vector normal to $\Si_t$, and $N$ be the outward unit normal vector of $S(t,u)$ that is tangential to $\Si_t$. With $T$ and $N$, we set the associated \emph{null frame} to be $(e_1,e_2,e_3,e_4)$. Here $(e_4,e_3):=(T+N,T-N)$ and $(e_1,e_2)$ is an orthonormal frame on $S(t,u)$. With $A,B\in\{1,2\}$ and $\D$ being the covariant derivative, we further introduce the null decomposition of the Ricci coefficients.
\begin{align*}
\chib_{AB}&:=\g(\D_A e_3, e_B),\qquad\quad\; \chi_{AB}:=\g(\D_A e_4, e_B),\\
\trchb&:=\de^{AB}\chib_{AB},\qquad\qquad\;\;\;\;\hchb_{AB}:=\chib_{AB}-\frac{1}{2}\trchb\,\g_{AB},\\
\trch&:=\de^{AB}\chi_{AB},\qquad\qquad\;\;\;\;\hch_{AB}:=\chi_{AB}-\frac{1}{2}\trch\,\g_{AB},\\
\omb&:=\frac 1 4 \g(\D_3e_3 ,e_4),\qquad\quad \,\,\,\;\,\; \om:=\frac 1 4 \g(\D_4 e_4, e_3), \\
\etab_A&:=\frac 1 2 \g(\D_4 e_3, e_A),\qquad\quad\;\;  \eta_A:=\frac 1 2 \g(\D_3 e_4, e_A),\\
\ze_A&:=\frac 1 2 \g(\D_{e_A}e_4, e_3),\qquad\quad\, \xib_A:=\frac 1 2 \g(\D_3 e_3,e_A),
\end{align*}
and curvature components
\begin{align*}
\a_{AB}&:=\R(e_A, e_4, e_B, e_4),\quad \;\;\,\aa_{AB}:=\R(e_A, e_3, e_B, e_3), \\
\b_{A}&:=\frac 1 2 \R(e_A, e_4, e_3, e_4), \;\;\quad \;\bb_{A}:=\frac 1 2 \R(e_A, e_3, e_3, e_4),\\
\rho&:=\frac 1 4 \R(e_3, e_4, e_3, e_4), \quad\;\;\;\;\;\; \rhod:=\frac{1}{4}{^*\R}(e_3,e_4,e_3,e_4),
\end{align*}
where $^*\R$ denotes the Hodge dual of Riemann tensor $\R$. We also denote $\nab$ as the induced covariant derivative on $S(t,u)$ and let $\nab_3\psi$ and $\nab_4\psi$ represent the projections of $\D_3\psi$ and $\D_4\psi$ to $S(t,u)$. 

We remark that the final angular momentum $a_f$ and final mass $m_f$ of the spacetime are determined by taking the limit of geometrically constructed parameters associated to a family of finite admissible \textit{General Covariant Modulated} (GCM) spacetimes. This construction is anchored on a GCM sphere $S_*$. Klainerman-Szeftel \cite{KS:Kerr2} introduced the associated angular momentum of $S_*$ as
\begin{align}\label{dfKS}
J:=r^5(\curl\b)_{\ell=1}.
\end{align}
Here $(\c)_{\ell=1}$ denotes the projection onto $\ell=1$ modes on the sphere. In their proof of Kerr stability, the definition of $J$ and its evolution play a crucial role. The intrinsic geometry of $S_*$ is uniquely determined by the uniformization theorem and its extrinsic properties are fixed by GCM conditions. For other definitions of angular momentum, interested readers are referred to \cite{CWY,CWWY,Rizzi}.

\vspace{2mm}

\emph{Stability of Minkowski---}The global nonlinear stability of Minkowski spacetime for the Einstein vacuum equations has been first established by Christodoulou-Klainerman \cite{Ch-Kl} in 1993. In 2007, Bieri \cite{Bieri} provided an important extension that requires one less derivative and weaker decay requirement for initial data compared to \cite{Ch-Kl}. In \cite{Shenglobal}, the third author extended the results of \cite{Bieri} to minimal decay assumptions, as stated in Theorem \ref{globalresult} below. The proofs presented in \cite{Ch-Kl,Bieri,Shenglobal} are based on the maximal-null foliation. In 2003, Klainerman-Nicol\`o \cite{Kl-Ni} revisited Minkowski stability in the exterior region of an outgoing null cone using the double null foliation. Later on, the third author \cite{ShenMink} reproved the stability of Minkowski in this exterior region with more general initial data. For the latest updates and more related details about stability of Minkowski, interested readers are referred to \cite{ShenMink,Shenglobal} and references therein.
\begin{thm}[Global stability of Minkowski \cite{Shenglobal}]\label{globalresult}
Let $s\in(1,2]$ and consider a \emph{$s$--asymptotically flat} initial data set $(\Si_0,g,k)$, i.e.,
\begin{align*}
g_{ij}=\de_{ij}+o(r^{-\frac{s-1}{2}}),\quad k_{ij}=o(r^{-\frac{s+1}{2}}), \quad i,j\in\{1,2,3\}.
\end{align*}
Then, the nonlinear stability of Minkowski holds true in the future of $\Si_0$. Moreover, we have the following asymptotic behaviors in the exterior region:
\begin{align*}
\Gag=O(r^{-\frac{s+1}{2}}),\qquad \Gab,\,\Gaw=O(r^{-1}\ujp^{-\frac{s-1}{2}}),
\end{align*}
where
\begin{align*}
\Gag&:=\left\{r\nab\trch,\,r\nab\trchb,\,\hch,\,\om,\,\omb,\,\ze,\,\etab,\,r\a,\,r\b,\,r\rho,\,r\rhod\right\},\\
\Gab&:=\left\{\eta,\,\trch-\frac{2}{r},\,\trchb+\frac{2}{r},\,\xib\right\},\;\;\Gaw:=\left\{\hchb,\,r\bb,\, u\aa\right\}.
\end{align*}
\end{thm}
In this paper, we report that we can improve estimates in Theorem \ref{globalresult} as below. These improvements of the extra decay rates in $r$ enable us to demonstrate the weighted asymptotic limits of various geometric quantities. In particular, with these limits, we establish the formula for the angular momentum memory effect.
\begin{lem}\label{globalresultimproved}
Under the same assumptions as in Theorem \ref{globalresult}, we have the following improved decay estimates
\begin{align}
\Gab&=O(r^{-\frac{s+1}{2}}).\label{GabsameasGag}
\end{align}
\end{lem}
\begin{proof}
To derive decay estimates in \eqref{GabsameasGag}, it suffices to construct a maximal-null foliation from a solved last slice $\Si_*$ and integrate $\Gab$ along $e_4$ backwardly from $\Si_*$.
\end{proof}
We summarize these improved hyperbolic estimates as
\begin{thm}\label{globalresulttoapply}
Let $s\in(1,3)\cup (3, 4)$ and require the initial data set $(\Si_0,g,k)$ to be $s$--asymptotically flat. 
Then, the spacetimes arising from these initial data are associated with the following asymptotic behaviors:
\begin{align}
\begin{split}\label{improvedestimates}
\Gag,\,\Gab&=O\left(r^{-\frac{s+1}{2}}+r^{-\frac{3}{2}}\right),\;\;\Gaw=O(r^{-1}\ujp^{-\frac{s-1}{2}}),\\
\Gar&=O\left(r^{-\frac{s+1-\de}{2}}+r^{-2}|u|^{\frac{3-s}{2}}\right),
\end{split}
\end{align}
where $\Gar:=\Gag\cup\Gab\setminus\{\eta,\,\xib,\,\omb\}$ and $0<\de\ll|s-3|$.
\end{thm}
\begin{proof}
Theorem \ref{globalresult} and Lemma \ref{globalresultimproved} directly imply that \eqref{improvedestimates} holds in the case $s\in(1,2]$. The estimates for $\Gag$, $\Gab$ and $\Gaw$ in \eqref{improvedestimates} in the cases $s\in(2,3)$ and $s\in (3, 4)$ follow from \cite{Bieri} and \cite{ShenMink}, respectively. Next, utilizing the incoming null structure equations and Bianchi equations, we derive
\begin{align*}
    \nab_3\Gar=r^{-1}\Gaw=\begin{cases}
        &O\left(r^{-2}|u|^{-\frac{s-1}{2}}\right),\qquad\quad s\in(3,4), \\
        &O\left(r^{-\frac{s+1-\de}{2}}|u|^{-1+\frac{\de}{2}}\right),\quad s\in(1,3).
    \end{cases}
\end{align*}
Integrating it by $u$, we then obtain the estimate for $\Gar$ in \eqref{improvedestimates} as stated.
\end{proof}
\emph{Stability of Kerr.---}The Kerr black hole is a $2$--parameter family of solutions $(K(a,M),\g_{a,M})$ discovered by Kerr \cite{Kerr} that solve \eqref{EVE}. In Boyer-Lindquist coordinates, its metric $g_{a,M}$ takes the form of
\begin{align}
\begin{split}\label{BL}
 \g_{a,M}=&-\frac{(\De-a^2\sin^2\th)}{q^2}dt^2-\frac{4aMr}{q^2}\sin^2\th dt d\varphi\\
 &+\frac{q^2}{\De}dr^2+q^2d\th^2+\frac{\Si^2}{q^2}\sin^2\th d\vphi^2
\end{split}
\end{align}
with $q^2:=r^2+a^2\cos^2\th$, $\quad\De:=r^2+a^2-2Mr$ and $\quad\Si^2:=(r^2+a^2)^2-a^2\sin^2{\th}\De$.

The nonlinear stability of Kerr black holes remains open for a long time, until a breakthrough emerged recently with the confirmation of Kerr stability for small angular momentum. This is due to a series of works by Klainerman-Szeftel, Giorgi-Klainerman-Szeftel and the third author \cite{KS:Kerr1,KS:Kerr2,KS:main,GKS,Shen}. The main results can be stated as follows:
\begin{thm}[Kerr stability with small angular momentum \cite{KS:main}]\label{kerrstab}
Let $(\Si_0,g,k)$ be a perturbed initial data set of a Kerr metric $\g_{a,M}$ with $|a|\ll M$.
The future globally hyperbolic development of $(\Si_0,g,k)$ has a complete future null infinity $\II_+$ and converges in its causal past $J^-(\II_+)$ to another nearby Kerr solution $\g_{a_f,M_f}$. Moreover, in the region $r\geq|u|^{1+\de}$, the following decay estimates hold 
\begin{align}
\begin{split}\label{Kerrdecay}
\trchc,\,\trchbc,\,\hch,\,\ze,\,\etab,\,\eta,\,\xi&=O(r^{-2}|u|^{-\frac{1}{2}-\de}),\\
\xib,\,\hchb,\,\omb,\,r\bb,\,\aa&=O(r^{-1}|u|^{-1-\de}),\\
\rho,\,\si&=O(r^{-3}),\\
\a,\,\b&=O(r^{-\frac{7}{2}-\de}).
\end{split}
\end{align}
Here $\trchc:=\trch-\frac{2}{r}$ and $\trchbc:=\trchb+\frac{2}{r}\left(1-\frac{2M_f}{r}\right)$.
\end{thm}
\emph{Main equations.---}Here we list the equations that we work with. For tensor fields defined on a $2$--sphere $S$, we denote by $\sk_0$ the set of pairs of scalar functions, $\sk_1$ the set of $1$--forms and $\sk_2$ the set of symmetric traceless $2$--tensors.
\begin{df}
For a given $\xi\in\sk_1$, we define
\begin{align*}
\sdiv\xi:= \de^{AB}\nab_A\xi_B,\qquad\curl\xi:= \in^{AB}\nab_A\xi_B.
\end{align*}
The Hodge operators are also denoted as
\begin{align*}
d_1 \xi &:=(\sdiv\xi,\curl\xi),\qquad\qquad\qquad\,\xi\in\sk_1,\\
(d_2 U)_A &:=\nab^BU_{AB},\qquad\qquad\qquad\quad\;\;\,\, U\in\sk_2, \\
d_1^*(f,f_*)_{ A}&:=-\nab_Af+\in_{AB}\nab_B f_*,\;\;\;\, (f,f_*)\in\sk_0.
\end{align*}
\end{df}
For later use, we list following equations based on the null-frame decompositions of Einstein vacuum equations.
\begin{prp}\label{nullstructure}
The Codazzi and torsion equations read
\begin{align}
\sdiv\hch&=\frac{1}{2}\nab\trch+\frac{1}{2}\trch\,\ze-\b+\hch\c\ze,\label{Codazzihch}\\
\curls\ze&=\rhod-\frac{1}{2}\hch\wedge\hchb,\label{torsion}
\end{align}
where $\hch\wedge\hchb:=\in^{AB}\hch_{A}{}^C\,\hchb_{CB}$. We also have
\begin{align}
\begin{split}
\nab_3\b+\trchb\,\b=&-d_1^*(\rho,-\rhod)+2\omb\c\b+\xib\c\a\\
&+2\hch\c\bb+3(\eta\rho+{^*\eta}\rhod).\label{Bianchieq}   
\end{split}
\end{align}
\end{prp}
\begin{proof}
See Propositions 7.3.2 and 7.4.1 in \cite{Ch-Kl}.
\end{proof}
\emph{Future null infinity.---}We denote the future null infinity by $\II_+$. Based on the hyperbolic estimates in \cite{Shenglobal} and \cite{KS:main}, in the same fashion as to \cite{BieriMemory}, we have
\begin{prop}\label{dfnullinfinity}
The following limits exist along $\II_+$:
\begin{align*}
\Ab&:=\lim_{C_u,r\to\infty}r\aa,\qquad\qquad\quad\;\Bb:=\lim_{C_u,r\to\infty}r^2\bb,\\
(\The_3,Z_3)&:=\lim_{C_u,r\to\infty}\nab_3(r^2\hch,r^2\ze),\quad\Thb:=\lim_{C_u,r\to\infty}r\hchb,\\
(\PP_3,\QQ_3)&:=\lim_{C_u,r\to\infty} \nab_3\left(r^3(\rho,\rhod)-\frac{r^3}{2}(\hch\c\hchb,\hch\wedge\hchb)\right).
\end{align*}
\end{prop}
We further define
\begin{equation}\label{limitofJ}
    (\PP,\QQ,\The,Z):=\int_{u} (\PP_3,\QQ_3,\The_3,Z_3)\, du.
\end{equation}
For any quantity $X$ defined along $\II_+$, we also denote
\begin{align*}
    X^+:=\lim_{u\to+\infty}X(u,\c ),\qquad X^-:=\lim_{u\to-\infty}X(u,\c).
\end{align*}
Multiply by an appropriate weight $r^p$ and letting $r\to \infty$ along $C_{u}$, we deduce the following lemma.
\begin{lem}\label{nullinfinityGa}
The following equations hold along $\II_+$:
\begin{align}
\The_3&=-\Thb,\label{Sieq}\\
\divo\Thb&=\Bb,\label{Codazzihchb}\\
(\PP_3,\QQ_3)&=-\dodo_1\Bb+\frac{1}{2}\left(|\Thb|^2,0\right),\label{renorequation}\\
\curlo Z&=\QQ, \label{curldivThe}
\end{align}
with $\nabo:=\lim_{C_u,r\to\infty} r\nab$.
\end{lem}
\emph{Displacement memory effect.---}We state the displacement nonlinear memory effect, which was established by Christodoulou in \cite{Ch} and Blanchet-Damour in \cite{BD}, extended by Bieri in \cite{BieriMemory} to slow decaying scenario with $s=2$, and further extended here in the setting of our Theorem \ref{globalresulttoapply} with $s\in(1,3)$.
\begin{thm}\label{memoryeffectprp}
The difference $\The^+-\The^-$ is uniquely determined by the elliptic system
\begin{align}\label{Chrmemoryeffect}
\dodo_1\dodo_2(\The^+-\The^-)&=(\PP^+-\PP^--4F,\,\QQ^+-\QQ^-),
\end{align}
where $F(\c):=\frac{1}{8}\int_{u}|\Thb(u,\c)|^2 du$. Furthermore, it holds
\begin{equation}\label{displace}
    \The^+-\The^-=O(\ujp^\frac{3-s}{2}).
\end{equation}
\end{thm}
\begin{proof}
It follows from \eqref{Sieq}, \eqref{Codazzihchb} and \eqref{renorequation} that
\begin{align*}
    \dodo_1\dodo_2\The_3=(\PP_3,\QQ_3)-\frac{1}{2}\left(|\Thb|^2, 0\right).
\end{align*}
Integrating it by $u$, we then deduce \eqref{Chrmemoryeffect} as stated.

To estimate the size of $\The^+-\The^-$, from Theorem \ref{globalresulttoapply} and Proposition \ref{dfnullinfinity} we have
    \begin{align*}
        \Thb=O(\ujp^{-\frac{s-1}{2}}).
    \end{align*}
Applying \eqref{Sieq}, we infer
    \begin{align*}
        \The^+-\The^-=\int_{u}O(\ujp^{-\frac{s-1}{2}})\, du=O(\ujp^\frac{3-s}{2}).
    \end{align*}
This concludes the proof of Theorem \ref{memoryeffectprp}.
\end{proof}
The difference $\The^+-\The^-$ is proportional to the permanent displacement of test masses in a gravitational wave detector. See \cite{Ch} for more details. Theorem \ref{memoryeffectprp} reveals that for $s\in (1, 3)$, this memory effect grows as $|u|^{\f{3-s}{2}}$ and it is caused by the growth of the so-called electric memory $\PP^+-\PP^-$ and magnetic memory $\QQ^+-\QQ^-$. These two memories are introduced by Bieri in \cite{BieriMemory} with $s=2$. One is also referred to \cite{BieriGarfinkle} for broader discussions. Theorem \ref{memoryeffectprp} further generalizes her result to the range $s\in (1, 3)$.

\vspace{2mm}

\emph{A new angular momentum memory.---}In this paper, we also investigate the evolution of the angular momentum defined in \cite{KS:Kerr2} along the future null infinity $\II_+$ based on hyperbolic estimates established in Theorem \ref{globalresulttoapply} with $s\in (3, 4)$ and Theorem \ref{kerrstab}. We find and report a new formula for the change of angular momentum along $\II_+$, which is named as the \textit{angular momentum memory effect}.
\begin{thm}\label{conservationlaw}
Along $\II_+$, the following limit  exists 
\begin{equation*}
    \JJ_3:=\lim_{C_u,r\to\infty}\nab_3J.
\end{equation*}
Moreover, we have
\begin{align}\label{conservationlaweq}
\JJ_3=(\The\wedge\Thb)_{\ell=1}+2\left(\curlo(\The\c\divo\Thb)\right)_{\ell=1}.
\end{align}
Furthermore, letting $\JJ:=\int_u \JJ_3\,du$, we have
\begin{equation}\label{angularmemoryeqn}
\JJ^+-\JJ^-=\int_{u}\left(\The\wedge\Thb+2\,\curlo(\The\c\divo\Thb)\right)_{\ell=1} du.
\end{equation}
\end{thm}
\begin{proof}
Differentiating the Codazzi equation \eqref{Codazzihch} by $r\curl$ and projecting it onto $\ell=1$ modes, we obtain
\begin{align*}
r(\curl\sdiv\hch)_{\ell=1}+r(\curl\b)_{\ell=1}=(\curl\ze)_{\ell=1}+\Gar\c\Gar.
\end{align*}
Noticing that $(d_1d_2U)_{\ell=1}=0$ for any $U\in\sk_2$, we deduce
\begin{equation}\label{betazeta}
    r(\curl\b)_{\ell=1}=(\curl\ze)_{\ell=1}+\Gar\c\Gar.
\end{equation}
By virtue of properties of $\ell=1$ modes, for any scalar function $h$, we have
\begin{align*}
    \left((r^2\De+2)h\right)_{\ell=1}=0.
\end{align*}
Hence, it follows from \eqref{torsion} and \eqref{betazeta} that
\begin{align*}
r^2(\De\rhod)_{\ell=1}&=-2(\rhod)_{\ell=1}+\left((r^2\De+2)\rhod\right)_{\ell=1}\\
&=-2(\curl\ze)_{\ell=1}-(\hch\wedge\hchb)_{\ell=1}\\
&=-(2r\curl\b+\hch\wedge\hchb)_{\ell=1}+\Gar\c\Gar.
\end{align*}
Together with 
\eqref{dfKS}, this implies
\begin{align}\label{siell=1}
    (r^5\De\rhod+r^3\hch\wedge\hchb)_{\ell=1}=\trchb\, J+r^3\,\Gar\c\Gar.
\end{align}
Next, differentiating \eqref{Bianchieq} by $r^5\curl$, we deduce
\begin{align*}
&r^4\nab_3(r\curl\b)+\trchb(r^5\curl\b)\\
=&-r^5\curl d_1^*(\rho,-\rhod)+2r^5\curl(\hch\c\bb)+r^3\,\Gar\c\Gab\\
=&-r^5\De\rhod+2r^5\curl(\hch\c\bb)+r^3\,\Gar\c\Gab.
\end{align*}
Combining with \eqref{dfKS} and \eqref{siell=1}, we then obtain
\begin{align*}
    &r^4\nab_3(r\curl\b)_{\ell=1}+\trchb\,J\\
    =&-r^5(\De\rhod)_{\ell=1}+2r^5(\curl(\hch\c\bb))_{\ell=1}+r^3\,\Gar\c\Gab\\
    =&r^3(\hch\wedge\hchb)_{\ell=1}-\trchb\,J+2r^5(\curl(\hch\c\bb))_{\ell=1}+r^3\,\Gar\c\Gab.
\end{align*}
Applying \eqref{dfKS} and \eqref{improvedestimates}, we derive for $s\in (3, 4)$
\begin{align}
\nab_3J&=r^4\nab_3(r\curl\b)_{\ell=1}+2\trchb\, J+r^3\,\Gar\c\Gab \nonumber\\
&=r^3(\hch\wedge\hchb)_{\ell=1}+2r^5(\curl(\hch\c\bb))_{\ell=1} \nonumber\\
&+r^3\, O\left(r^{-\frac{s+1-\de}{2}}\right)\, O\left(r^{-\frac{3}{2}}\right). \label{nab 3 J eqn asymptotics}
\end{align}
Taking $r\to\infty$ and employing Proposition \ref{dfnullinfinity}, we conclude
\begin{align*}
\JJ_3=(\The\wedge\Thb)_{\ell=1}+2\left(\curlo\left(\The\c\Bb\right)\right)_{\ell=1}.
\end{align*}
The desired equality \eqref{conservationlaweq} follows by inserting \eqref{Codazzihchb}. Integrating \eqref{conservationlaweq} in $u$, we present the formula of the angular momentum memory \eqref{angularmemoryeqn}.
\end{proof}
Notice that when $s\in (2, 3)$, if $r^2 \hch$ tends to $\The$ at $\II_+$, then \eqref{conservationlaweq} still holds. 
The hyperbolic estimates in Theorem \ref{globalresulttoapply} and Theorem \ref{kerrstab} further enable us to measure the size of $\JJ^+-\JJ^-$.
\begin{thm}\label{angularmemory}
In both settings of Theorem \ref{globalresulttoapply} with $3<s<4$ and Theorem \ref{kerrstab}, it holds
\begin{equation}
    \JJ^+-\JJ^-=O(1).
\end{equation}
\end{thm}
\begin{proof}
Employing \eqref{improvedestimates} and \eqref{displace}, we deduce
\begin{align*}
\left(\The\wedge\Thb+2\,\curlo(\The\c\divo\Thb)\right)_{\ell=1}&= O(\ujp^{\frac{1-s}{2}})\, O(\ujp^{\frac{3-s}{2}})\\
&=O(\ujp^{2-s}).
\end{align*}
Plugging it in \eqref{angularmemoryeqn}, we have for $s\in(3 ,4)$
\begin{align*}
    \JJ^+-\JJ^-=\int_u O(\ujp^{2-s})\, du=O(1).
\end{align*}
In the stability of Kerr regime, by virtue of \eqref{Kerrdecay} in Theorem \ref{kerrstab}, we conclude
\begin{align*}
\JJ^+-\JJ^-=
\int_u O(|u|^{-\frac{3}{2}-2\de})\, du=O(1)
\end{align*}
as stated.
\end{proof}
Consequently, Theorem \ref{angularmemory} indicates that $\JJ^+-\JJ^-$ can be precisely measured in perturbed Minkowski spacetimes with $s\in (3, 4)$ and in the context of Kerr stability with small angular momentum \cite{KS:main}. However, when $s\in (2, 3)$, the term on the right of \eqref{conservationlaweq} is not integrable, thus $\JJ^+$ diverges.
\vspace{2mm}

\indent\emph{Newman-Penrose formalism.---}We proceed to rewrite nonlinear displacement memory \eqref{Chrmemoryeffect} and our new angular momentum memory \eqref{angularmemoryeqn} in NP formalism. This formalism has been first introduced in \cite{NP}. With the following tetrad:
\begin{align*}
l:=e_4,\;\;\;\; n:=\frac{1}{2}e_3,\;\;\;\; m:=\frac{e_1+ie_2}{\sqrt{2}},\;\;\;\;\mm:=\frac{e_1-ie_2}{\sqrt{2}},
\end{align*}
we define the spin coefficients as below:
\begin{align}
\begin{split}\label{NPGamma}
{}^{\text{(NP)}}\si &:=-\g(\D_m l,m),\qquad  {}^{\text{(NP)}}\la:=\g(\D_\mm n,\mm),\\
{}^{\text{(NP)}}\b&:=-\frac{1}{2}\left(\g(\D_m l,n)-\g(\D_m m,\mm)\right),\\
{}^{\text{(NP)}}\a&:=-\frac{1}{2}\left(\g(\D_\mm l,n)-\g(\D_\mm m,\mm)\right).
\end{split}
\end{align}
We also denote the Weyl-NP scalar as
\begin{equation}\label{NPR}
    {}^{\text{(NP)}}\Psi_2:=\R(l,m,\mm,n).
\end{equation}
By definition, we can express ${}^{\text{(NP)}}\si, {}^{\text{(NP)}}\la, {}^{\text{(NP)}}\Psi_2$ in terms of the Ricci coefficients and curvature components defined before as
\begin{align}
\begin{split}\label{NPidentities}
    &{}^{\text{(NP)}}\si=-(\hch_{11}+i\hch_{12}),\qquad {}^{\text{(NP)}}\la=\frac{1}{2}(\hchb_{11}-i\hchb_{12}),\\
    &{}^{\text{(NP)}}\Psi_2+{}^{\text{(NP)}}\si {}^{\text{(NP)}}\la=\f12\Big(\rho-i\rhod+\f12 (\hch \cdot \hchb-i\hch\wedge \hchb)\Big).
\end{split}
\end{align}
Taking into account the asymptotic behaviors of $\hch, \hchb$ and $\rho, \rhod$, we set
\begin{align}
\begin{split}\label{defX}
\Xh&:=\lim_{C_u,r\to\infty} r^2 \, {}^{\text{(NP)}}\si,\qquad \Xhb:=\lim_{C_u,r\to\infty} r \, {}^{\text{(NP)}}\la,\\
Y&:=\lim_{C_u,r\to\infty} r^3 \left({}^{\text{(NP)}}\Psi_2+{}^{\text{(NP)}}\si {}^{\text{(NP)}}\la\right).
\end{split}
\end{align}
We denote the angular derivatives $\de:=\nab_m$ and $\deb:=\nab_\mm$. It is also convenient to introduce the below corresponding spin derivatives for $k=1, 2$:
\begin{equation*}
    \de_k=\de+k\left({}^{\text{(NP)}}\b-{}^{\text{(NP)}}\ov{\a}\right), \;\;\deb_k=\deb+k\left({}^{\text{(NP)}}\ov{\b}-{}^{\text{(NP)}}\a\right).
\end{equation*}
With the assistance of spin derivatives, we can rewrite the Hodge operators $d_1$ and $d_2$ in the complex form:
\begin{lem}\label{Lemma de k}
Given $\xi\in\sk_1 $ and $U\in\sk_2$, we have
\begin{align*}
&\deb_1 (\xi_1+i \xi_2)=\f{\sdiv \xi+i\curl \xi}{\s2},\\
&\deb_2 (U_{11}+i U_{12})=\f{(\sdiv U)_1+i(\sdiv U)_2}{\s2}.
\end{align*}
\end{lem}
With these preparations, we now reformulate the memory effects \eqref{Chrmemoryeffect} and \eqref{angularmemoryeqn} in NP formalism.
\begin{prp}\label{NPtheorem}
 Along the future null infinity $\II_+$, the Christodoulou's memory effect shows
\begin{align}\label{Chris memory NP}
\debo_1\debo_2 (\Xh^+-\Xh^-)&=-\left(\ov{Y}^+-\ov{Y}^-\right)+2\int_{-\infty}^{+\infty}|\Xhb|^2. \end{align}
Our newly introduced angular momentum effect reads
\begin{align}\label{angular momentum memory NP}
\JJ^+-\JJ^-&=4\int_{-\infty}^{+\infty}\im\left(\Xh\,\Xhb-2\debo_1(\Xh\deo_2\Xhb)\right)_{\ell=1}.
\end{align}
Here the symbol $^\circ$ is defined in Lemma \ref{nullinfinityGa}.
\end{prp}
\begin{proof}
From \eqref{NPidentities} and \eqref{defX}, we have
\begin{align*}
\Xh&=-(\The_{11}+i\The_{12}),\qquad\Xhb=\frac{1}{2}(\Thb_{11}-i\Thb_{12}),\\
Y&=\f12(\PP-i\QQ).
\end{align*}
In view of Lemma \ref{Lemma de k}, a directly computation yields
\begin{align*}
\debo_1 \debo_2 \Xh=&-\f{1}{\s2} \debo_1\left( (\divo \The)_1+i(\divo \The)_2\right)\\=&-\f{1}{2}\left(\divo\divo \The+i\,\curlo\divo\The \right).
\end{align*}
Plugging this into \eqref{Chrmemoryeffect}, we obtain
\begin{align*}
\debo_1 \debo_2 (\Xh^+-\Xh^-)=&-\f{1}{2}\big(\PP^+-\PP^--4F+i(\QQ^+-\QQ^-) \big)\\=&-(\ov{Y}^+-\ov{Y}^-)+2\int_{-\infty}^{+\infty}|\Xhb|^2
\end{align*}
as stated. To prove \eqref{angular momentum memory NP} for the angular momentum memory, we first observe that
\begin{align}\label{Im Xh Xhb}
\im(\Xh\,\Xhb)&=-\f12(-\The_{11}\Thb_{12}+\The_{12}\Thb_{11})=\f{1}{4}\The\wedge\Thb.
\end{align}
Applying Lemma \ref{Lemma de k} again, we then deduce
\begin{align*}
\Xh\left(\deo_2\Xhb\right)&=-\f{1}{2\s2}(\The_{11}+i\The_{12})\left((\divo \Thb)_1-i (\divo\Thb)_2\right)\\
&=-\f{1}{2\s2}\left((\The\c\divo\Thb)_1+i(\The\c\divo\Thb)_2\right),
\end{align*}
which renders
\begin{equation}\label{de 1 Xh de2 Xhb}
\im\left(\debo_1(\Xh\,\deo_2\Xhb)\right)=-\f{1}{4}\curlo(\The\c\divo\Thb).
\end{equation}
Combining \eqref{Im Xh Xhb} and \eqref{de 1 Xh de2 Xhb}, we infer
\begin{align*}
\im\left(\Xh\,\Xhb-2\debo_1(\Xh\deo_2\Xhb)\right)
=\f{1}{4}\The\wedge\Thb+\f{1}{2}\curlo\left(\The\c\divo\Thb\right).
\end{align*}
Injecting the above expression in \eqref{angularmemoryeqn}, we then finish the proof of Proposition \ref{NPtheorem}.
\end{proof}
\indent\emph{Physical discussions.---}This part is devoted to the detection procedure of the memory effects in the laboratory. Gravitational memory effect can be measured by the interferometric gravitational wave detectors \cite{Ashtekar, Ch}. To measure Christodoulou's memory effect, we consider two test masses $m_1$ and $m_2$ sitting initially at a right angle observed from a reference mass $m_0$ with distances $d_0$.  We assign $\Si_t$ with the normal coordinates $(x_1, x_2, x_3)$ spread by the exponential map $\exp(x_1 e_1+x_2 e_2+x_3 N)$ starting from $m_0$. Suppose that the source of gravitational wave is in the direction of $N$. Thus, the horizontal plane $(x^1,x^2)$ is tangent to $S_{t ,u}$. Denote $(x^i_{(B)})_{i=1,2,3}$ to be the spatial coordinates of the test mass $m_B$ with $B=1,2$ and set the initial positions and initial velocities as $(x^i_{(B)})^-=d_0\de_B^i$ and $(\dot{x}^i_{(B)})^-=0$. According to \cite{Ch} (and (5) in \cite{Rizzi}), we have
\begin{align}
\begin{split}\label{xdotx}
    (x^A_{(B)})^+-(x^A_{(B)})^-&=-\frac{d_0}{r}(\The^+_{AB}-\The^-_{AB}),\\
    \dot{x}^A_{(B)}&=\frac{d_0}{2r}\Thb_{AB}.
\end{split}
\end{align}
\indent The spin memory effect was introduced by Pasterski-Strominger-Zhiboedov in \cite{PSZ}. We calculate this effect in our setup and propose a possible device. 

\begin{figure}[ht]
\includegraphics[width=8.6cm]{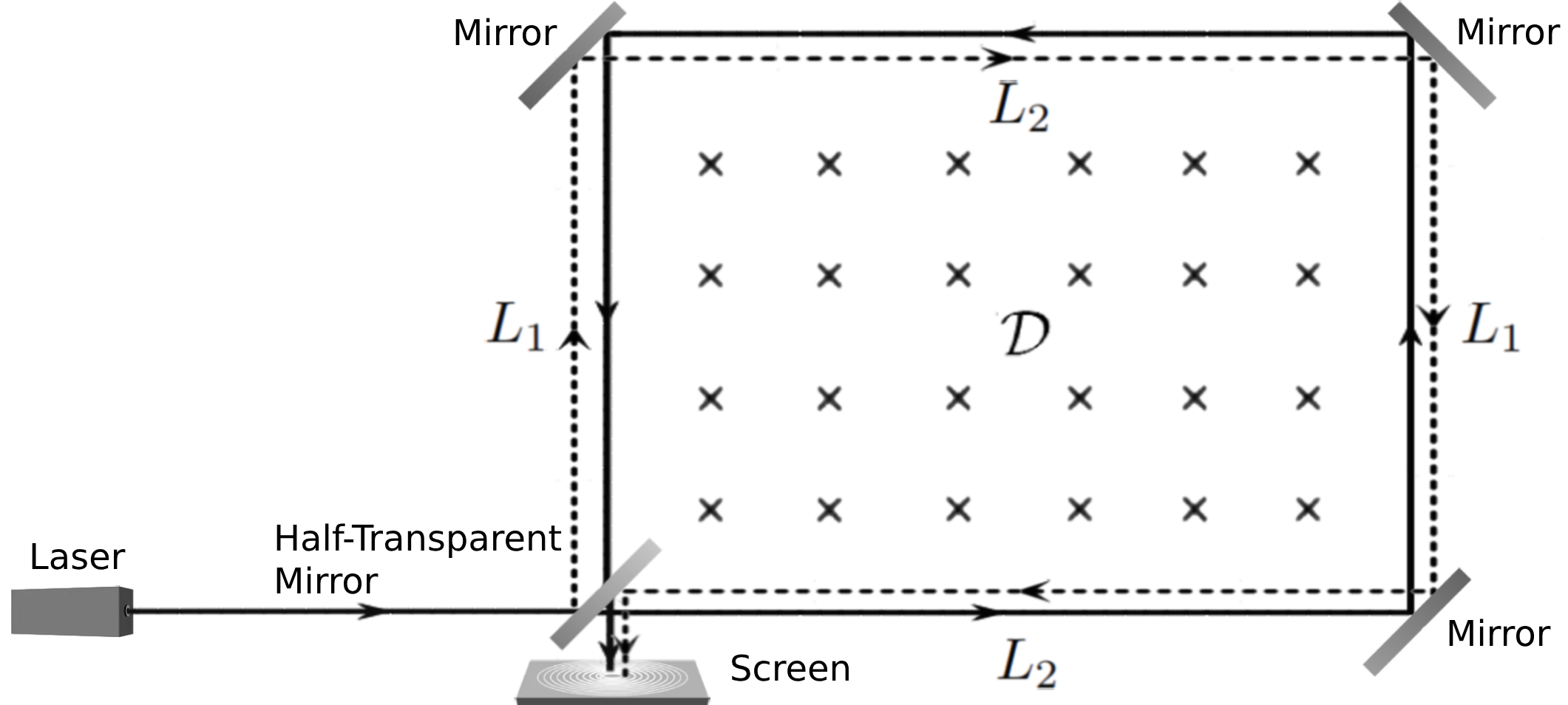}
\caption{Measurement of spin memory effect}
\label{spinfigure}
\end{figure}

The spin memory effect in \cite{PSZ} is designed to be detected by experimental equipment placed on a sphere with fixed area radius $r$. We note that the experiment can also be carried out with a laser source, a half-transparent mirror, 3 mirrors placed as in Figure \ref{spinfigure}. We denote $\DD$ to be a rectangular region with lengths $L_1,L_2\ll r$. By placing a screen detector as in Figure \ref{spinfigure}, the total time delay $\De\tau$ between the clockwise ray and the counterclockwise ray traveling around the loop $\pr\DD$ can be measured.
Proceeding as in (4.5) of \cite{PSZ}, we obtain
\begin{equation}\label{PSZspin}
\De\tau=\frac{1}{|\pr\DD|}\int_u du\oint_{\pr\DD}Z_A\,dx^A.
\end{equation}
Consequently, applying Green's theorem, it follows
\begin{align*}
\oint_{\pr\DD}Z_A\,dx^A&=\iint_{\DD}\curl Z\, dx^1\wedge dx^2\\
&=r^{-1}\iint_{\DD}\QQ\, dx^1\wedge dx^2,
\end{align*}
where we used
\eqref{curldivThe}. Hence, back to \eqref{PSZspin}, we deduce
\begin{equation}\label{Detaueq}
\De\tau=\frac{1}{r|\pr\DD|}\int_u du\iint_{\DD}\QQ\, dx^1\wedge dx^2.
\end{equation}

Recall that, from \eqref{limitofJ} we have
\begin{equation*}
 \QQ=\int_{u} \lim_{C_u,r\to\infty} \nab_3 \left(r^3\rhod\right) du-\frac{1}{2}\The\wedge\Thb.
\end{equation*}
Compared with \eqref{angularmemoryeqn}, we can see that part of our angular momentum memory is reflected in the spin memory effect. The formula \eqref{Detaueq} also has an explanation rooted in the classic electromagnetic theory. The quantity $\QQ$ represents the magnetic part of the memory effect. According to Faraday's Law, when a closed loop is placed in a changing magnetic field, an electric current will be induced in the loop, which causes the time delay.

\vspace{2mm}

\indent\emph{Acknowledgements.---}The authors would like to thank Sergiu Klainerman and J\'er\'emie Szeftel for many helpful discussions and remarks. The authors would also like to thank Jiandong Zhang for a valuable discussion on spin memory.

\bibliographystyle{apsrev4-1}

\end{document}